\documentclass[runningheads]{llncs}
\usepackage{amssymb}
\usepackage{multirow,setspace,verbatim,amsfonts,amsmath,amsbsy,epsfig,url}
\usepackage{gensymb}
\usepackage{booktabs}

\usepackage{cite}
\usepackage{makecell}
\usepackage{epstopdf}
\usepackage{array,multirow}
\newcolumntype{P}[1]{>{\centering\arraybackslash}p{#1}}
\newcolumntype{M}[1]{>{\centering\arraybackslash}m{#1}}

\usepackage{algorithm}
\usepackage[noend]{algpseudocode}
\usepackage{tikz}
\usepackage{mathrsfs} 
\usepackage[algo2e]{algorithm2e} 
\usepackage{array}
\usepackage{color}
\usepackage{graphicx}
\begin{document}
\title{Severity Quantification and Lesion Localization of COVID-19 on CXR using Vision Transformer}
%
%
\author{Gwanghyun Kim\inst{1}\and
Sangjoon Park\inst{1}\and
Yujin Oh\inst{1}\and
Joon Beom Seo\inst{2}\and
Sang Min Lee\inst{2}\and
Jin Hwan Kim\inst{3}\and
Sungjun Moon\inst{4}\and
Jae-Kwang Lim\inst{5}\and
Jong Chul Ye\inst{1}}
\authorrunning{G. Kim et al.}
%
\institute{Korea Advanced Institute of Science and Technology, Daejeon, South Korea \\
\email{\{gwang.kim, depecher, yujin.oh, jong.ye\}@kaist.ac.kr}\and
{Asan Medical Center, University of Ulsan College of Medicine, Seoul, South Korea} \\ \and
{College of Medicine, Chungnam National Univerity, Daejeon, South Korea} \\ \and
{College of Medicine, Yeungnam University, Daegu, South Korea} \\ \and
{School of Medicine, Kyungpook National University, Daegu, South Korea}}

\maketitle              

\begin{abstract}

Under the global pandemic of COVID-19, building an automated framework that quantifies the severity of COVID-19 and localizes the relevant lesion on chest X-ray images has become increasingly important. Although pixel-level lesion severity labels, e.g. lesion segmentation, can be the most excellent target to build a robust model,  collecting enough data with such labels is difficult due to time and labor-intensive annotation tasks. Instead,  array-based severity labeling that assigns integer scores on six subdivisions of lungs can be an alternative choice enabling the quick labeling. Several groups proposed deep learning algorithms that quantify the severity of COVID-19 using the array-based COVID-19 labels and localize the lesions with explainability maps. 
To further improve the accuracy and interpretability, here we propose a novel Vision Transformer tailored for both quantification of the severity and clinically applicable localization of the COVID-19 related lesions. Our model is trained in a weakly-supervised manner to generate the full probability maps from weak array-based labels. Furthermore, a novel progressive self-training method enables us to build a model with a small labeled dataset. The quantitative and qualitative analysis on the external testset demonstrates that our method shows comparable performance with radiologists for both tasks with stability in a real-world application.

\end{abstract}
\section{Introduction}
The ongoing coronavirus disease 2019 (COVID-19) pandemic has resulted in 115,028,175 confirmed cases with 2,551,329 death cases worldwide as of March 2,  2021 \cite{ng2020covid}. As  pneumonia is commonly
present in COVID-19 patients, radiological examinations are often used for the diagnosis of COIVD-19 \cite{shi2020radiological}.
In particular,
chest X-ray (CXR) has comparative advantages in terms of the short scan time, low cost, and a low dose of radiation over chest computed tomography (CT) \cite{jacobi2020portable}. Therefore, there is a great potential to use CXR to analyze the patient's condition, such as severity quantification or lesion localization. In particular,
a deep learning-based algorithm that quantifies the severity and localizes COVID-19 lesions on CXR image may help radiologists under global pandemic. 

Although pixel-level segmentation labels have the most abundant information toward this goal, it is hard to collect a large dataset due to its time-consuming annotation. To mitigate this issue, simple array-based severity labeling methods are introduced, where integer-valued severity scores are assigned on the six or eight subdivisions of CXR images \cite{toussie2020clinical, borghesi2020covid}. With the array labels, several algorithms \cite{sig2020covid, cohen2020predicting} quantify the severity of COVID-19  and generate explainability maps using convolutional neural networks and visualization methods such as  GradCAM \cite{selvaraju2017grad}, LIME \cite{ribeiro2016should}, etc.  However, 
the probability values on the explainability maps, usually based on the normalized activation \cite{selvaraju2017grad}, are not directly related to the real probability value of the lesion existence.
 Accordingly,  comparisons of the saliency maps with the true lesion annotation from the radiologists are rarely made.

To provide clinically meaningful quantification of severity and localization of  COVID-19 lesion, here we propose a novel Vision Transformer (ViT) trained in a weakly-supervised manner using severity array labels.
Recently, Vision Transformer (ViT) \cite{dosovitskiy2020image} was shown to attain state-of-the-art (SOTA) performance on the image classification tasks by learning long-range dependency among pixels using a self-attention mechanism \cite{vaswani2017attention}. Training a vanilla ViT requires a vast dataset to learn inductive biases, so that the authors of \cite{dosovitskiy2020image} suggest using a hybrid ViT that uses a convolutional neural network (CNN) as a feature embedding network on the small-sized dataset. 
By extending the idea, our vision Transformer is trained using the low-level CXR feature corpus that are
generated using a feature extraction network pretrained on a large CXR dataset. 
Additionally,  we use ROI max-pooling layer that can bridge between pixel-level supervision and the severity array label in a weakly-supervised manner \cite{oquab2015object}.

One of the important advantages of our novel ViT scheme for severity quantification and lesion localization is that the global attention maps from Transformer
can lead to full lesion maps where each pixel value directly means the probability of the abnormality of COVID-19. Moreover,
 our novel progressive self-training, which was inspired by \cite{xie2020self},
 enables to utilize the large unlabeled dataset in addition to the small severity-labeled dataset. 

By performing both quantitative and qualitative evaluation using the external test data set,
we validate the model's performance and its generalization capability for different institute data set.


\section{Method}

\begin{figure}[t!]
\centering
\setlength{\intextsep}{5pt plus 1pt minus 1pt}
\includegraphics[width=12cm]{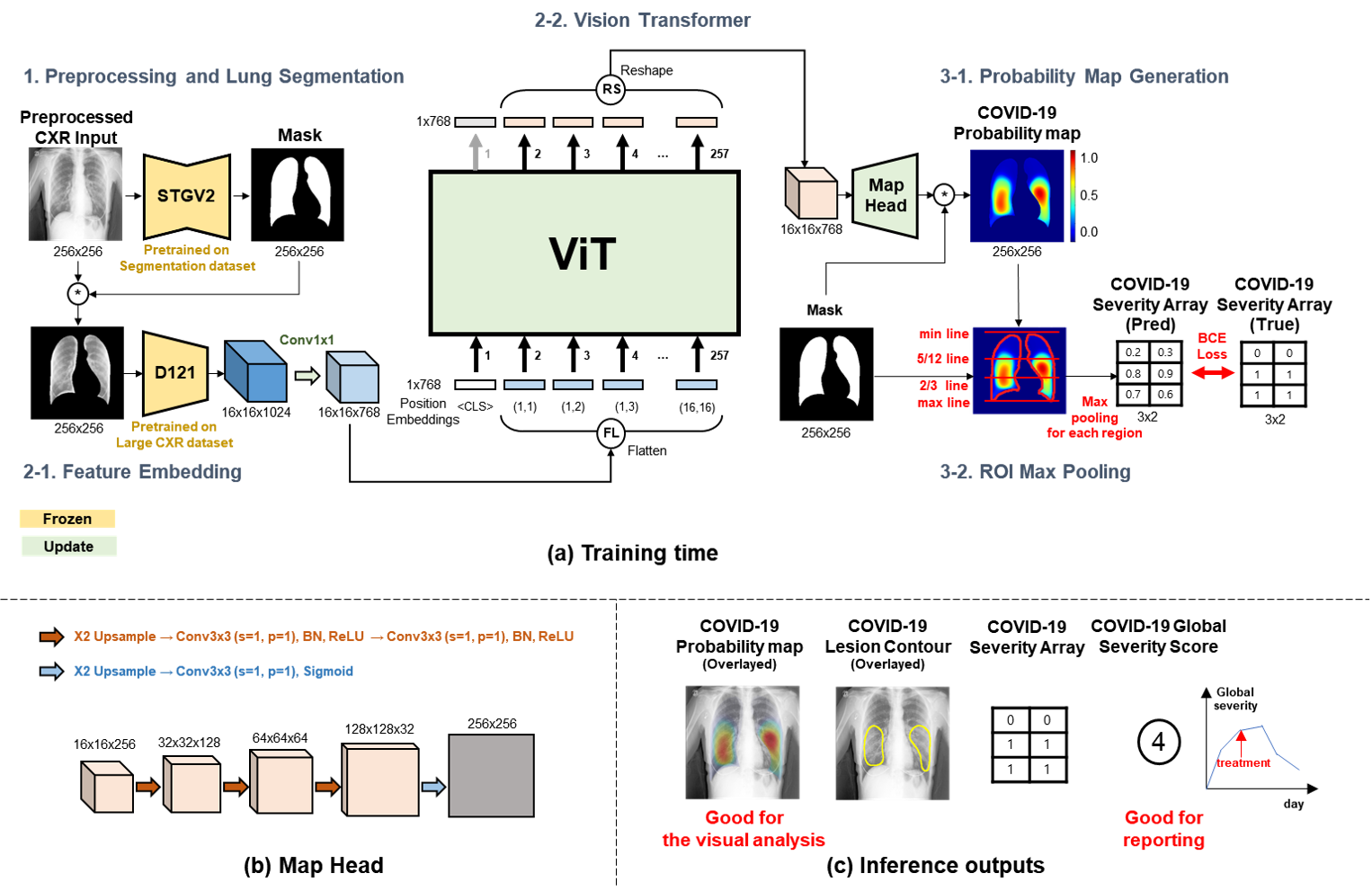}
\caption{ Overview of the proposed framework. }
\label{fig:fig1}
\end{figure}

\subsection{Severity annotation method}
Our model is trained with the frontal CXR images with the severity score arrays annotated following the method in \cite{toussie2020clinical}. Specifically,
each lung is first subdivided into three areas in the vertical direction. The lower area extends from the intercostal groove to the lower hilar mark. The middle area extends from the lower hilar mark to the upper hilar mark, and the upper area extends from the upper hilar mark to the tip. 
Then, each area is divided into two regions along the horizontal direction across spines.
Binary score 0/1 is assigned to each region according to the absence/presence of the opacity \cite{toussie2020clinical}. 
Accordingly, the completed label has a form of 3$\times$2 array and a global severity score that is the sum of all elements ranges 0-6.

\subsection{Proposed Model}
The proposed model's overall architecture is illustrated in Fig. \ref{fig:fig1} (a). Firstly, an input CXR image is preprocessed and given to the lung segmentation network. The segmented lung image is fed into the feature embedding network, followed by a Vision Transformer. The generated final features from Vision Transformer are provided into the map head that generates the full COVID-19 probability map. By ROI max pooling, the 3$\times$2 COVID-19 severity arrays are estimated as the final output of our model. 


\subsubsection{Hybrid Vision Transformer Backbone.}
For the feature embedding network to extract a low-level CXR feature corpus, the network of \cite{ye2020weakly} that won first place in CheXpert  \cite{irvin2019chexpert} challenge is employed, which applies probabilistic class activation map (PCAM) pooling to the output from the DenseNet-121 based feature extractor to enhance localization performance as well as classification performance. The feature extractor is pretrained on the vast public CXR image dataset \cite{irvin2019chexpert} classifying 10 radiological findings: pneumonia, consolidation, lung opacity, pleural effusion, cardiomegaly, edema, atelectasis, pneumothorax, support devices, and no finding. 
Specifically, we use 16$\times$16$\times$1024 features before transition layer 3 of DenseNet-121.

Specifically, the segmented lung $\boldsymbol{x} \in \mathbb{R}^{{H}\times{W}\times{C}}$ is projected into the feature $\boldsymbol{c} \in \mathbb{R}^{{H'}\times{W'}\times{C'}}$ through the feature embedding network $\boldsymbol{F}$. The feature vector $c$ $\in \mathbb{R}^{C'}$ is the projected vector at each pixel location.

\begin{flalign}
&&\boldsymbol{c} &= \boldsymbol{F}(\boldsymbol{x}), & \boldsymbol{x} \in \mathbb{R}^{{H}\times{W}\times{C}}, \; \boldsymbol{c} \in \mathbb{R}^{{H'}\times{W'}\times{C'}}\\
&&\boldsymbol{c} &= [c^{1};c^{2};c^{3};...;c^{{H'}\times{W'}}], & c \in \mathbb{R}^{{C'}}
\end{flalign}

For Vision Transformer, we adopt the ViT-B/16 architecture of \cite{dosovitskiy2020image}, which input is $1024$$\times$$16$$\times$$16$ patches. We first embed projected features ${c}  \in \mathbb{R}^{{C'}}$  into $ {c}_{p} \in \mathbb{R}^{{D}}$  using $1$ $\times$ $1$ convolution kernel. Learnable vector $c_\texttt{cls}\;$  that is the role of \texttt{[class]} token of BERT \cite{devlin2018bert} is included for the training of our ViT. However, we utilize the final layer outputs of the ViT-B/16 except for the output at the token position.    Also, positional embedding $\textbf{E}_{pos}$ is added not to lose the positional information of the feature map. Multihead self-attention (MSA), multi-layer perceptron (MLP), layer normalization (LN), and residual connections in each block which are essential parts of the standard transformer \cite{devlin2018bert} are used equally in our model.
Formally, this procedure can be written by
\begin{flalign}
&&[c_{p}^{1};c_{p}^{2};c_{p}^{3};...;c_{p}^{{H'}\times{W'}}] &= \texttt{conv}([c^{1};c^{2};c^{3}; ...;c^{{H'}\times{W'}}]), &c_{p} \in \mathbb{R}^{{D}}\\
&&[z^{0}_0;z^{1}_0;z^{2}_0;...;z^{{H'}\times{W'}}_0] &= [c_\texttt{cls};c_{p}^{1};c_{p}^{2};c_{p}^{3};...;c_{p}^{{H'}\times{W'}}] + \textbf{E}_{pos}\\
&&\textbf{z}_0 &= [z^{0}_0;z^{1}_0;z^{2}_0;...;z^{{H'}\times{W'}}_0]\\
&&\textbf{z}'_{l} &= \texttt{MSA}(\texttt{LN}(\textbf{z}_{l-1})) + \textbf{z}_{l-1}, &l = 1 ... L\\
&&\textbf{z}_{l} &= \texttt{MLP}(\texttt{LN}(\textbf{z}'_{l})) + \textbf{z}'_{l}, &l = 1 ... L
\end{flalign}
where $L$ denotes the number of layers in ViT (i.e. $L=12$ for the case of ViT-B/16).

\subsubsection{Probability Map Generation and ROI Max Pooling.}
A map head using the output of ViT is composed of 4 upsizing convolutional blocks
and generates a map which size is the same as the input size. The detailed architectures of the map head are illustrated in Fig. \ref{fig:fig1} (b). Multiplying the output of the map head with the lung mask $m \in \mathbb{R}^{{H}\times{W}}$, the COVID-19 lesion probability map $y \in \mathbb{R}^{{H}\times{W}}$ is generated.
ROI max-pooling (RMP) is used for converting the COVID-19 lesion map into the severity array $a \in \mathbb{R}^{{3}\times{2}}$ as depicted in Fig. \ref{fig:fig1}(a).

\begin{flalign}
&&y &= \texttt{MAPHEAD}([z^{1}_{L};z^{2}_{L};z^{3}_{L};...;z^{{H'}\times{W'}}_{L}]) \otimes m, &y, m \in \mathbb{R}^{{H}\times{W}} \\
&&a &= \texttt{RMP}(y, m), &a \in \mathbb{R}^{{3}\times{2}}
\end{flalign}

Specifically, the lungs are separated into the right and left lung by computing the lung mask's connected components. Next, the lines that split each lung into three zones are estimated at 5/12 and 2/3 locations of the line between the highest and lowest lung mask location. Then, the max value of each subdivision is assigned to each corresponding element of the 3$\times$2 array. For optimizing the model, a binary cross-entropy loss is calculated between the predicted severity array and the label severity array. These line estimation and max-pooling processes are the keys to a weakly supervised learning scheme. We evaluate the weakly-supervised learning scheme's validity by performing both quantitative and qualitative external tests in Section 3.

\subsection{Progressive Self-training for the Unlabeled Dataset}
\begin{figure}[hbt!]
\centering
\includegraphics[width=11cm]{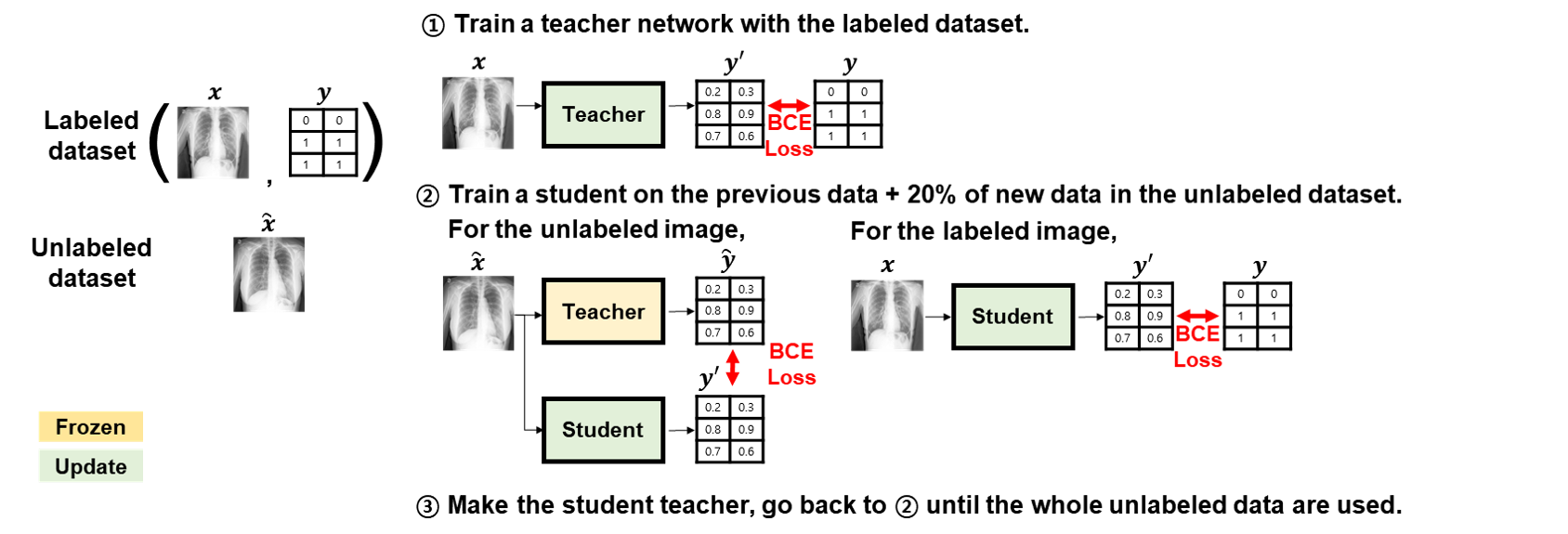}
\caption{The illustration of progressive self-training method.}
\label{fig:fig2}
\end{figure}

Under the pandemic situation,  it is often difficult to collect enough severity labels even if the labeling method is quite simple.
Motivated by \cite{xie2020self}, 
 we employ the progressive self-training that utilizes the larger severity-unlabeled dataset, as well as the small severity-labeled dataset, so that it can improve the performance of the model. The detail procedure of the self-training method is in Fig. \ref{fig:fig2}. First, a teacher network is trained with the labeled dataset. In the second step, a new student network, a copy of the teacher, is trained on the previous dataset combined with new data in a subset of the unlabeled dataset. The model is optimized with the labeled input's true label and with the pseudo labeled generated from the teacher network for the unlabeled input. Next, the student becomes a new teacher, and the process is iterated, going back to the second step.

\section{Experimental details}

\begin{table}[b!]
\caption{Summary of main datasets for training and test of our framework.}
\label{table:tab1}
\centering
\scalebox{0.7}{
\begin{tabular}{M{2.5cm}|M{2cm}|M{2cm}|M{2cm}|M{2.5cm}}
\hline
\multicolumn{1}{c|}{Sources} & Total & Train  & Validation &  External Test \\ \hline
YNU                          & 286   & 226    & 60         & -           \\
KNUH                         & 293   & 226    & 67         & -           \\
Brixia \cite{sig2020covid}   & 4,413 & 3,335  & 978        & -         \\
CNUH                         & 81    & -      & -          & 81            \\
BIMCV \cite{de2020bimcv}     & 12    & -      & -          & 12            \\ \hline
Total                        & 4,985 & 3,787  & 1,105      & 93            \\ \hline
\end{tabular}}
\end{table}

\subsubsection{Dataset.}
For the segmentation network training,  a total of 247 normal CXR images from the publicly available JSRT dataset \cite{shiraishi2000development} with its segmentation label from the publicly available SCR dataset \cite{van2006segmentation} are used. 
To provide reliable segmentation masks even for abnormal CXR,
the segmentation network is then fine-tuned 
 in a semi-supervised manner using a total of 680 CXR images of pneumonia, tuberculosis was used from public archives \cite{cohen2020covid}.
 
For the pretraining of the feature embedding network to generate low-level CXR feature corpus,
we used 190,847 frontal CXR images from the publicly available CheXpert dataset \cite{irvin2019chexpert} where the presence of 14 types of radiological findings are annotated (no finding, enlarged cardiom., cardiomegaly, lung lesion, lung opacity, edema, consolidation, pneumonia, atelectasis, pneumothorax, pleural effusion, pleural other, fracture, supported devices).
For training of ViT and the map head, the domestic dataset from 2 independent institutions (Yeungnam University Hospital [YNU], Daegu, Korea; Kyungpook National University Hospital [KNUH], Daegu, Korea) is combined with the publicly available Brixia dataset \cite{sig2020covid}. Two board certificated radiologists labeled severity with consensus following the labeling method in \cite{toussie2020clinical} for the domestic dataset. Brixia labeling \cite{sig2020covid, borghesi2020covid} is different from the method of \cite{toussie2020clinical} in terms of severity score scale (0-3) for each subdivision and the anatomic landmarks determining the split line. However, the authors of \cite{sig2020covid} mentioned that this severity label could be extended to the labeling \cite{toussie2020clinical} that we use by mapping the score in $\{$1,2,3$\}$ to 1 for each zone with slight differences, which procedure we followed. 

For external testing of the quantification and the localization performance of the model,  CXR images from another independent domestic institution, Chungnam  National Univerity Hospital [CNUH], are used for the quantitative evaluation. The severity labels are annotated from the same two radiologists who labeled CXR images from the other domestic training dataset. In the publicly available BIMCV dataset  \cite{de2020bimcv}, the COVID-19 lesion segmentation label is annotated on 12 frontal images. We use these images for the qualitative analysis of our model.  The numbers of images for quantifying severity and the localization of COVID-19 lesion are summarized in Table \ref{table:tab1}.

\subsubsection{Implementation details.}
Our preprocessing method contains resizing to 256$\times$ 256, Gaussian blurring with a 3$\times$ 3 kernel, histogram equalization, and normalization. 
For our segmentation model, we use
Adam optimizer \cite{kingma2014adam} with a learning rate of 0.0001, batch size of 1 for the training of 20,000 steps.
For the feature extraction network, Adam optimizer \cite{kingma2014adam} with a learning rate of 0.0001, batch size of 8, and binary cross-entropy loss are used for binary classification of 10 pathologies for 160,000 steps of training. 
For ViT and the map head training, the weights of the segmentation and feature extraction network are frozen, and SGD optimizer with momentum of 0.9 and learning rate of 0.004, batch size of 8 is used for 12,000 steps. The models at 12,000, 11,500 and 11,000 steps are ensembled for the test. We adopt  Mean Squared Error (MSE) as a main metric for the regression of the global severity score ranging 0-6,  but also calculated Mean Absolute Error (MAE), Correlation Coefficient (CC), $R^2$  score for the global score regression, and mean of area under the ROC curve (AUC) for the binary classification of each region in severity array. 
The proposed networks are implemented with PyTorch framework and trained using an NVIDIA V100.

\begin{table}[b!]
\caption{The severity prediction results on CNUH external testset of the proposed frameworks with different backbones. }
\label{table:tab2}
\centering
\scalebox{0.7}{
\begin{tabular}{M{4cm}|M{1.9cm}|M{1.1cm}|M{1.1cm}|M{1.1cm}|M{1.1cm}|M{1.9cm}}
\hline
Backbone      & Params (M) & \multicolumn{1}{c|}{MSE} & \multicolumn{1}{c|}{MAE} & \multicolumn{1}{c|}{CC} & \multicolumn{1}{c|}{$R^2$} & \multicolumn{1}{c}{Mean AUC} \\ \hline
D121+ViT-B/16 & 95.5        & \textbf{1.889}                    & 0.926                    & \textbf{0.760}                              & \textbf{0.520}                      & \textbf{0.882}                        \\
ResNet-18 \cite{he2016deep}           & 3.4         & 2.012                    & 0.975                    & 0.735                              & 0.488                      & 0.875                        \\
DenseNet-121 \cite{huang2017densely}          & 13.6        & 2.000                    & \textbf{0.914}                    & 0.730                              & 0.491                      & 0.873                        \\
DenseNet-201 \cite{huang2017densely}         & 38.0        & 2.395                    & 1.086                    & 0.676                              & 0.391                      & 0.849                        \\
ResNet-152 \cite{he2016deep}          & 52.5        & 2.235                    & 0.951                    & 0.716                              & 0.432                      & 0.856                        \\ 
NASNet-Large \cite{zoph2018learning}      & 56.1        & 2.592                    & 0.963                    & 0.715                              & 0.341                      & 0.873                        \\ \hline
\end{tabular}}
\end{table}

\begin{table}[b!]
\caption{The severity prediction results on CNUH external testset of the proposed frameworks varying training dataset setting. }
\label{table:tab3}
\centering
\scalebox{0.7}{
\begin{tabular}{M{4.8cm}|M{1.1cm}|M{1.1cm}|M{1.1cm}|M{1.1cm}|M{1.9cm}}
\hline
Brixia dataset for training   & MSE   & MAE   & CC    & $R^2$ & Mean AUC \\ \hline
not included                  & 1.926 & 1.012 & 0.744 & 0.510 & 0.808    \\
supervised                    & 1.889 & 0.926 & 0.760 & 0.520 & 0.882    \\
semi-supervised (progressive) & \textbf{1.296} & \textbf{0.827} & \textbf{0.836} & \textbf{0.670} & \textbf{0.892}    \\
semi-supervised (one step)    & 1.963 & 1.074 & 0.729 & 0.501 & 0.848    \\ \hline
\end{tabular}}
\end{table}

\begin{figure}[b!]
\centering
\includegraphics[width=8cm]{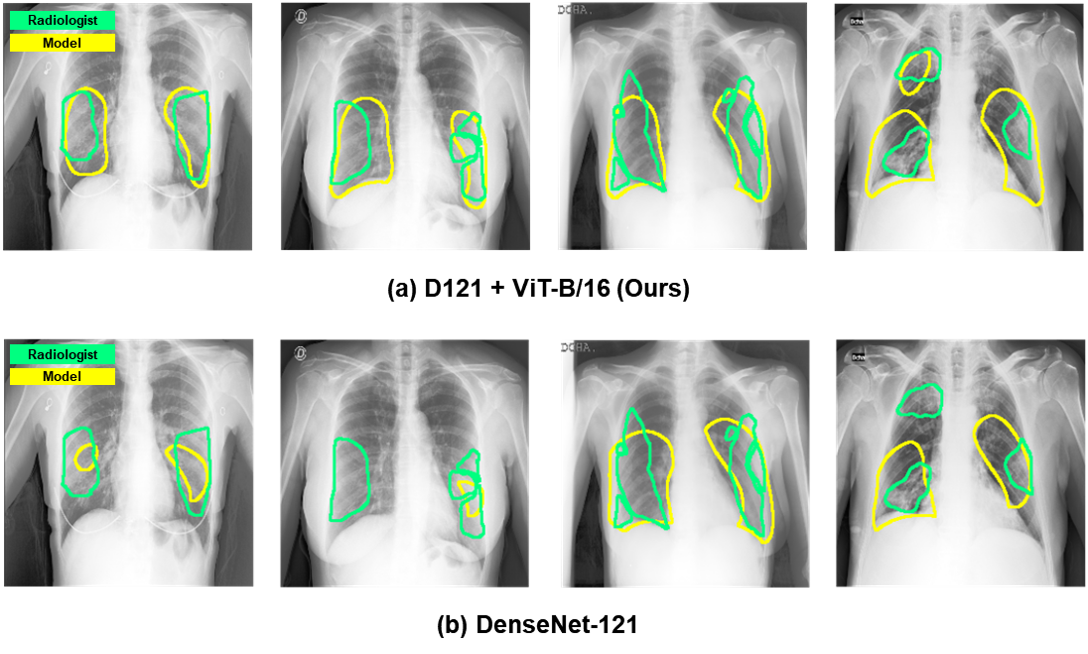}
\caption{The examples of localization results on the images in BIMCV \cite{de2020bimcv} dataset of the proposed framework with the proposed ViT backbone and DenseNet-121 \cite{huang2017densely} backbone. Green: the expert radiologists' annotation. Yellow: the model's prediction obtained by thresholding the probability map of 0.5.}
\label{fig:fig3}
\end{figure}

\section{Experimental Results}

\subsubsection{Comparison of our ViT based model with CNN-based models.}
To validate the superiority of ViT based architecture of our model (D121+ViT-B/16) over CNN-based architectures, we compare the performance of severity quantification and localization of COVID-19 lesions with those of various CNN models: ResNet-18, ResNet-152 \cite{he2016deep}, DenseNet-121, DenseNet-201 \cite{huang2017densely}, NasNet-Large \cite{zoph2018learning}. All models except ours were trained based on ImageNet \cite{deng2009imagenet} pretrained weights, and the same training setting from our proposed model was applied.

The quantitative comparison of severity quantification performance on CNUH external testset is represented in Table \ref{table:tab2}. Our model outperformed the CNN-based models for most of the metrics, demonstrating superior performance and generalizability.
We also perform the qualitative comparison of the localization performance on BIMCV \cite{de2020bimcv} external testset between our ViT based model and DenseNet-121 \cite{huang2017densely} based model. The result is represented in Fig. \ref{fig:fig3}. Our proposed model's prediction of the abnormal region on the CXR image shows more accurate localization than DenseNet-121 \cite{huang2017densely} based model.

\subsubsection{Performance of progressive self-training.}
To evaluate the usefulness of the progressive self-training methods, we compare the models trained with four different training data sets depending on the utilization of Brixia dataset, as described in Table \ref{table:tab3}.  In default (not included), the models are trained only on the domestic training dataset. For the third model proposed, the inclusion of Brixia data without labels increased progressively after every 2,000 of the total 12,000 steps as described in Fig. \ref{fig:fig2}. For the last model, after 2,000 steps of training only on the domestic dataset and the whole Brixia data \cite{sig2020covid} without labels are included all at one, and another 10,000 steps of training is followed. External test results on CNUH testset are represented in Table \ref{table:tab3}. The model trained on the progressive self-training outperforms all methods, proving the effectiveness of the method and implying the labeling method discrepancy between the domestic dataset and Brixia dataset \cite{sig2020covid}. Brixia dataset \cite{sig2020covid} provides a consensus subset where five different radiologists annotate 150 CXR images, and 
the MSE between the gold standard from the majority voting and each radiologist's global score is calculated to 1.683. Therefore,  the MSE value of 1.296 in our model can be accepted as reasonable considering the MSE of the radiologists.

\section{Conclusion}
In this study, we presented a Vision Transformer tailored to quantify the severity and the localization of COVID-19 related lesions on  CXR. 
Our novel Vision Transformer using a low-level CXR feature corpus enabled us to encode long-range dependency between pixels, which is crucial for the localization of the lesion. Furthermore, our method enabled the generation of the full COVID-19 lesion map, in which pixel values directly represent the probability of abnormality. In addition, progressive self-training methods allowed the use of the small severity annotated dataset and the large unlabeled dataset. Our model provided the results that are the comparable performance to the expert radiologists on the external testset, validating its generalizability.

     

\bibliographystyle{splncs04}
\bibliography{ref_MICCAI2021_kgh}

\end{document}